# Stacking-Dependent Optical Properties in Bilayer WSe$_2$

Kathleen M. McCreary,[1] Madeleine Phillips,[1] Hsun-Jen Chuang,[2] Darshana Wickramaratne,[1] Matthew Rosenberger,[3] C. Stephen Hellberg[1] and Berend T. Jonker[1]

[1] *Naval Research Laboratory*, Washington, DC 20375
[2] *Nova Research, Inc.* Washington, DC 20375
[3] *University of Notre Dame*, Notre Dame IN 46556

Abstract

The twist angle between the monolayers in van der Waals heterostructures provides a new degree of freedom in tuning material properties. We compare the optical properties of WSe$_2$ homobilayers with 2H and 3R stacking using photoluminescence, Raman spectroscopy, and reflectance contrast measurements under ambient and cryogenic temperatures. Clear stacking-dependent differences are evident for all temperatures, with both photoluminescence and reflectance contrast spectra exhibiting a blue shift in spectral features in 2H compared to 3R bilayers. Density functional theory (DFT) calculations elucidate the source of the variations and the fundamental differences between 2H and 3R stackings. DFT finds larger energies for both A and B excitonic features in 2H than in 3R, consistent with experimental results. In both stacking geometries, the intensity of the dominant A$_{1g}$ Raman mode exhibits significant changes as a function of laser excitation wavelength. These variations in intensity are intimately linked to the stacking- and temperature-dependent optical absorption through resonant enhancement effects. The strongest enhancement is achieved when the laser excitation coincides with the C excitonic feature, leading to the largest Raman intensity under 514 nm excitation in 2H stacking and at 520 nm in 3R stacked WSe$_2$ bilayers.

Introduction

Two-dimensional transition metal dichalcogenides (TMDs) are an exciting class of materials that exhibit unique optical,[1,2] electronic,[3,4] and valleytronic[5–7] properties and have captivated researchers for the last decade. These layered materials are defined by the chemical formula $MX_2$, where M is a metal and X is a chalcogen. A TMD monolayer is actually three atomic layers thick, exhibiting a trigonal prismatic geometry where the outer chalcogen layers are covalently bonded to the central metal atom. Out of plane bonding between monolayers is governed by the relatively weak van der Waals forces. However, these interlayer interactions are strong enough that altering the stacking geometry can significantly modify the measured properties of TMD bilayers.

Several semiconducting TMD materials, including $WSe_2$, exhibit a transition from direct bandgap to indirect band gap upon increasing from one to two layers. Monolayers have received the bulk of attention thus far for applications ranging from chemical sensing[5] to quantum emission.[6] However, $WSe_2$ bilayer TMDs are particularly attractive for use in piezoelectric applications,[7] valleytronics[8], photodetectors,[9] and transistors,[10] in part due to their improved stability and higher carrier mobility. Additionally, bilayer systems exhibit an added degree of freedom: the twist angle. While early experimental works largely investigated the naturally occurring 2H stacking order[11,12] because of the reliance on mechanically exfoliated $WSe_2$, recent advances in synthesis[13–15] and sequential stacking processes[16] are opening the door to investigations of 3R stacked bilayers and twisted bilayer systems. The atomic positions in 2H and 3R $WSe_2$ are shown schematically in Fig. 1a and 1b. Recent optical studies noted differences between 2H and 3R stacked bilayers in the photoluminescence emission associated with direct

K-K transitions,[17] suggesting the orientation between layers may provide a way to controllably tune properties. However, the impact of stacking order on other optical properties was not addressed, and contradictory behaviors were later observed in other PL studies,[16] where PL emission energies of the K-K transition were the same for 2H and 3R stacking orders. Both studies employed sequential stacking to fabricate bilayers, which is susceptible to interlayer contaminants[18] as well as angle misalignments which could complicate the interpretation of data.

Here we provide a comprehensive investigation of the Raman, photoluminescence, and reflectance contrast response of 2H and 3R $WSe_2$ bilayers grown by chemical vapor deposition (CVD). The use of CVD ensures clean residue-free interlayer interfaces and precise stacking orientation which are difficult to insure in bilayers assembled by transfer processes. Clear stacking-dependent differences are evident in all optical properties in CVD-grown bilayers, and properties discussed in this work can serve as a point of reference for homobilayers assembled by mechanical transfer.

Results and discussion

$WSe_2$ bilayers are obtained via chemical vapor deposition at a growth temperature of 825 °C on $SiO_2$/Si substrates, similar to previously reported processes.[19] Solid $WO_3$ (1000 mg, Alfa Aesar 99.999%) and solid selenium (500 mg, Alfa Aesar 99.999%) serve as precursors for the growth. Additional details are provided in the Methods. The high growth temperatures along with differences in the thermal expansion coefficients between $SiO_2$ and transition metal dichalcogenides produces tensile strain in $WSe_2$ following synthesis.[20,21] To remove the strain, all bilayer samples are picked up from the growth substrate using a polymer stamp transfer process and subsequently placed on fresh $SiO_2$/Si.[22]

As is evident in Fig. 1c and d, $WSe_2$ islands containing both monolayer and bilayer regions are produced. Notably, the inner bilayer region and larger outer monolayer equilateral triangles exhibit one of two orientations, either at a 60 degree angle or rotationally aligned to one another. These two sample types correspond to the different stacking orders, 2H and 3R, respectively.[23] Imperfections such as wrinkles, cracks, or tears, can be introduced during the transfer process. However, regions that are free of macroscopic defects or deformations are readily available in both 2H and 3R bilayer stacks and typically exhibit lateral dimensions on the order of several microns to tens of microns, enabling reliable optical measurements.

The optical properties of representative 2H and 3R $WSe_2$ bilayers are investigated using a commercial Horiba LabRam Evolution confocal spectrometer integrated with a closed cycle Montana optical cryostat. A 50× objective focuses the desired illumination source to a spot having approximately 2 um diameter on the sample. Raman spectroscopy, photoluminescence (PL) measurements, and reflectance data are all collected at the same spot on each sample. 532 nm excitation is used for Raman and PL measurements, with laser power at the sample kept below 100 uW for all measurements. For reflectance measurements, a broad band white light source is focused through the objective onto the sample and the reflected intensity is measured. To minimize unwanted contributions from the substrate, we determine the reflectance contrast RC = $(R_{off}-R_{on})/R_{off}$, where $R_{on}$ is the reflectance of the $WSe_2$ sample on the $SiO_2$ substrate and $R_{off}$ is the reflectance of the bare $SiO_2$ substrate.

The room temperature Raman spectra of both 2H and 3R bilayer stacking orders exhibit a dominant $A_{1g}$ mode at 251 cm$^{-1}$, as well as a second, much weaker, feature at 259 cm$^{-1}$ associated with the 2LA(M) mode, consistent with previous reports for bilayer WSe$_2$.[24,25] However, in our direct comparisons between the stacking orders, clear differences are evident. The $A_{1g}$ intensity of 2H WSe$_2$ bilayer (blue line, Fig. 2a) is a factor of three larger than in the 3R bilayers (red line, Fig. 2a) under identical excitation conditions. The smaller $A_{1g}$ intensity present in 3R WSe$_2$ allows for the detection of the nearly degenerate $E_{2g}$ mode, as shown in the multi-component fittings of the normalized spectra (Fig. 2 b,c). Bilayers with 2H stacking are well fit by a single Lorentzian at 251.0 cm$^{-1}$ and a gaussian at 259.8 cm$^{-1}$, attributed to the $A_{1g}$ and second order 2LA(M) modes, respectively. However, bilayers of 3R stacking exhibit a low wavenumber shoulder, requiring two Lorentzian curves at 249.6 cm$^{-1}$ and 251.3 cm$^{-1}$ to adequately fit the spectra, as shown in Figure 2c, attributed to the $E_{2g}$ and $A_{1g}$ modes. These differences between the stacking orders are consistently observed across multiple samples (SI Fig. S1).

Both PL and reflectance contrast measurements reveal clear differences between the optical behavior of the 2H and 3R stacking orders. The PL spectra (Fig. 2d) suggest the presence of multiple emission features in both stacking orders, with higher energy features likely associated with direct-gap transitions and lower energy features indirect-gap transitions. In 3R WSe$_2$, the emission features at higher energy exhibit significantly stronger emission relative to the lower energy features. In contrast, all emission components exhibit similar strength for 2H samples, leading to fundamentally different PL line shapes between the 2H and 3R samples. In addition, the emission extends to higher energy for the 2H samples. The absolute intensity of PL emission exhibits some sample-to-sample variation, but the distinctly different line shapes are consistently observed between 2H and 3R (SI Fig. S2). Reflectance contrast measurements (Fig. 2

e) find a dip centered at 1.66 eV (1.64 eV) for 2H (3R) WSe$_2$ bilayers, consistent with PL extending to higher energies for 2H stacking. These features are associated with the direct-gap A exciton at the K point in the Brillouin zone. The small absorption coefficient of the indirect gap transition prevents its detection in reflectance contrast measurements and any features at higher energy are indistinguishable from the background level.

To further elucidate these observations, measurements are repeated under cryogenic conditions at 4K. As evident in Fig. 3a, the low temperature reflectance contrast spectra reveal three clear features at 1.72eV (1.71eV), 2.18eV (2.15eV), and 2.42eV (2.37eV) for 2H(3R) stacking orders. Following previous convention, we label these as A, B, and C features, respectively.[26] The A exciton originates from a hole in the highest valence band at K and an electron in the lowest conduction band at K whose spin and layer-localization matches the highest valence band. The B exciton originates from a hole in the second-highest valence band at K (split from the highest valence band by the spin orbit interaction) and an electron in the lowest conduction band at K with appropriate spin and layer. The spin-splitting in the conduction band is much smaller than that in the valence band. The C feature has previously been experimentally reported in WSe$_2$,[26,27] but the precise origin remains unclear. It has been suggested that the C feature either corresponds to an excited state of the A exciton,[11,28] or originates from a band nesting region where valence and conduction bands are nearly parallel and lead to numerous optical transitions that are nearly degenerate in energy.[27,29–31] Here, we find the A, B, and C features are systematically at higher energy in 2H materials than in 3R, with the difference monotonically increasing from 10 meV for A to 50 meV for C.

The corresponding photoluminescence at 4K (Fig. 3b) finds features in 2H WSe$_2$ that are blue-shifted relative to similar features in the 3R sample. In both stacking orders, two well-separated features are evident. The peak at 1.71 eV (1.70 eV) for 2H (3R) matches well with the A exciton energy observed in reflectance contrast, indicating this feature is associated with the direct-gap A exciton. The second feature for 2H (3R) is centered around 1.56 eV (1.52 eV) and associated with indirect transitions. Notably, for both PL and reflectance contrast, the general behaviors remain the same across the investigated temperature range, with features in 2H WSe$_2$ always blue-shifted relative to their 3R counterpart.

In contrast, temperature considerably changes the behavior observed in the Raman spectra (Fig. 3c). At 4K, the dominant A$_{1g}$ Raman mode at 252 cm$^{-1}$ is four times stronger in 3R WSe$_2$ than in 2H, a reversal from the room temperature behavior where this mode was three times stronger in the 2H (Fig 2a). To elucidate this behavior, additional measurements are acquired under a variety of excitation wavelengths between 488 nm and 588 nm (Fig. 4). All data are normalized to the Si Raman peak at 523 cm$^{-1}$ to allow for direct comparison of the A$_{1g}$ intensity[11] and all laser powers are kept below 100 uW, to prevent sample damage. It is immediately evident that the resulting A$_{1g}$ intensity depends strongly on the excitation wave length used. In both 2H and 3R systems, large intensities are achieved for laser excitation wavelength between 514 nm and 532 nm (highlighted by the dashed blue box in Fig. 4). Laser excitations well above or well below this range lead to significantly lower A$_{1g}$ intensity. Additionally, the conditions leading to maximum A$_{1g}$ intensity are different for the two stacking orders, with 2H exhibiting highest signal for 514 nm excitation while the maximum in 3R WSe$_2$ is achieved for 520 nm excitation based on the discrete wavelengths used here.

The detailed $A_{1g}$ intensity as a function of excitation wavelength is presented in Fig. 5a. Upon comparison with the reflectance contrast data replotted as a function of wavelength (Fig. 5b), it is apparent that the Raman intensity and the reflectance contrast follow the same trend. The strongest enhancement occurs when the excitation wavelength coincides with the C exciton in the reflectance contrast spectra. A second enhancement, although to a much lesser degree, is also present near the B exciton feature. Similar resonantly enhanced Raman has been reported in mechanically exfoliated materials having 2H stacking,[11,32] but has not been investigated in 3R stacking orders. The C exciton resonant enhancement of Raman also clarifies the reason the 2H shows stronger $A_{1g}$ mode at room temperature and 3R shows stronger $A_{1g}$ mode at low temperature. As evident in Fig. 5a, 532 nm excitation (dashed line) is considerably closer to resonance with the C exciton feature in 3R stacking than in 2H stacking, leading to larger Raman enhancement in 3R $WSe_2$ at low temperature. However, as temperature increases, red shifts are expected in all excitonic features. The 4 K (Fig. 3a) and room temperature (Fig. 2e) reflectance contrast spectra show the A exciton red-shifting by about 30 nm. A similar shift would lead to room temperature C exciton peak positions at approximately 542 nm in 2H and 553 nm in 3R, consistent with our observations of larger Raman intensity in 2H than 3R $WSe_2$ at room temperature with 532 nm excitation (Fig. 2a). Thus the intrinsic differences in Raman intensity between the 2H and 3R stacking (e.g. at 488 nm, Fig 4a, i) are strongly modulated by resonance effects when the excitation energy is near resonance with the temperature dependent absorption edges.

To provide insight into the intimate relationship between the changes in stacking on the reflectance contrast, PL, and Raman of bilayer $WSe_2$, we perform density functional theory (DFT) calculations (See the Methods section for calculation details). We computed band structures for both the 2H and 3R bilayer (Fig. 6 a,b). We focus on energy differences between the calculated band gaps for the 2H and 3R stacking when comparing to the experimental results. We note that 3R bilayers lack inversion symmetry and may be present in two distinct geometries: AB and BA. DFT calculations find very similar results for 3R AB and BA structures, as detailed in the supplementary information (Fig S3). We therefore only present results for 3R AB structure below. Since we expect the binding energies of the A, B and C excitons to be approximately the same for the 2H and 3R structures, comparing the difference between the experimental exciton energies in the two structures will cancel out the binding energies. This allows us to compare changes in the single particle band gaps of our first-principles calculations with our experiments.

We approximate the A exciton values in our bilayers by computing the "A" bandgaps in each layer of the bilayer of a given structure and taking the average. We follow the same procedure for the B excitons. We compute "A" band gap values of 1.402 eV for 2H and 1.393 eV for 3R. The 2H "A" band gap is 9 meV higher than the 3R "A" band gap, which agrees with the 2H A exciton being ~10 meV higher than the 3R A exciton as shown in both our reflectance contrast and PL measurements at 4K. We compute "B" band gap values of 1.822 eV for the 2H structure and 1.803 eV for the 3R structure, so the 2H "B" band gap is 19 meV larger than the 3R "B" band gap. This is in agreement with the 2H B exciton being 30 meV higher than the 3R B exciton in low-temperature reflectance contrast. The energy difference between the A and B exciton features measured in the 4K reflectance contrast experiment is 460 meV in the 2H structure and 440 meV in the 3R structure. This is also in good agreement with our calculated values of $(B-A)^{2H}$ = 420 meV and $(B-A)^{3R}$ = 410 meV. Although first-principles calculations that use the GGA functional are known to underestimate the band gaps of the semiconducting TMDs,[33] we find that differences in our computed band gaps are consistent with our experiments (Fig. 6c).

We now turn to understanding the spectral dependence of the Raman intensity and the differences due to 2H and 3R stacking. We calculate the zone center vibrational modes of 2H and 3R bilayer $WSe_2$ and the Raman cross section for the $A_{1g}$ and $E_g$ modes of the bilayer structure (see Methods).

For the 2H structure we find the $A_{1g}$ mode at 246 cm$^{-1}$ and the $E_{2g}$ mode at 242 cm$^{-1}$. For the 3R AB and BA structures we calculate the $A_{1g}$ mode to be 245 cm$^{-1}$ and the $E_{2g}$ mode at 242 cm$^{-1}$. The calculated vibrational frequencies of $A_{1g}$ and $E_{2g}$ modes for both stackings are near-degenerate and are close to the experimentally measured frequency of 251 cm$^{-1}$. We calculate the cross section, $S(\omega)$, which is directly proportional to the Raman intensity for the $A_{1g}$ and $E_{2g}$ modes of the 2H and 3R (AB and BA) structures. The magnitudes of $S(\omega)$ for the 3R AB and BA structures are similar, hence we only present results for the 3R AB structure. We find the magnitude of $S(\omega)$ for the $E_{2g}$ mode is more than an order of magnitude lower compared to $S(\omega)$ for the $A_{1g}$ mode for the 2H and 3R structures. Hence, we only focus on $S(\omega)$, of the $A_{1g}$ mode of the 2H and 3R (AB) structure, which is illustrated in Fig. 7(b). Two caveats are in place. First, to account for the underestimation of the band gap in our GGA calculations, the cross-sections are blue-shifted so that the onset in the GGA dielectric functions used to calculate $S(\omega)$ coincide with GW calculations of the A exciton of monolayer $WSe_2$ [34] (commonly referred to as a scissors shift). Secondly, since we do not include the role of excitons this precludes an investigation of the potential role of the excited states of excitons in resonant Raman processes. We note that the excited states of excitons in monolayer $WSe_2$ have low oscillator strengths compared to the A, B and C excitons of monolayer $WSe_2$[35] and hence, are unlikely to lead to pronounced Raman intensity under resonant excitation.

With this in mind, our calculations indicate that $S(\omega)$, which is proportional to the Raman intensity, is peaked at an energy that coincides with where we find the C exciton of bilayer $WSe_2$ and is suppressed for energies that correspond to the A and B exciton. Taken together, the agreement between experiment and our calculations of both the electronic band structure and Raman cross section supports our hypothesis that (i) the blue shift in the band gaps going from 3R to 2H stacking is due to stacking-derived changes in the electronic structure and (ii) the large spectral dependence of the Raman intensity of the $WSe_2$ mode at ~250 cm$^{-1}$ is due to resonant effects that are enhanced when excitation occurs at the C exciton of the 2H and 3R structures.

Conclusion:

In conclusion, the photoluminescence, Raman, and reflectance contrast features of bilayer $WSe_2$ were found to be strongly dependent on stacking angle, with clear differences evident between 2H and 3R $WSe_2$. Reflectance contrast measurements find a notable blue shift of the A, B and C excitonic features in 2H $WSe_2$ compared to the 3R stacking order. Density functional calculations confirm this arises from fundamental differences in the band structure. Subsequently, we find that reflectance contrast and Raman spectra are intimately linked through resonant Raman effects, where a strong enhancement in the dominant $A_{1g}$ mode occurs when the laser excitation coincides with the C exciton energy. The laser wavelength leading to maximum enhancement is stacking-dependent, due to the differing C exciton emission energy and band structure, with 2H-WSe$_2$ exhibiting maximum enhancement under 514 nm excitation and 3R-WSe$_2$ at 520 nm excitation at cryogenic temperatures. This work provides much needed insight into bilayer TMD systems.

Methods:

<u>CVD growth of WSe$_2$ bilayers.</u> Prior to use, all SiO$_2$/Si substrates are cleaned in acetone, IPA, and Piranha etch (H$_2$SO$_4$+H$_2$O$_2$) then thoroughly rinsed in DI water. At the center of the furnace is positioned a quartz boat containing ~1 gram of WO$_3$ powder. Two SiO$_2$/Si wafers are positioned face-down, directly above the oxide precursor. A separate quartz boat containing selenium powder is placed upstream, outside the furnace-heating zone. The upstream SiO$_2$/Si wafer contains perylene-3,4,9,10-tetracarboxylic acid tetrapotassium salt (PTAS) seeding molecules, while the downstream substrate is untreated. The hexagonal PTAS molecules are carried downstream to the untreated substrate and promote lateral growth of the monolayer WSe$_2$. Pure argon (65 sccm) is used as the furnace heats to the target temperature. Upon reaching the target temperature of 825 °C, 10 sccm H$_2$ is added to the Ar flow and maintained throughout the 10 minute soak and subsequent cooling to room temperature.

<u>Polymer Stamp Transfer</u>. Polydimethylsiloxane (PDMS) is made from a commercially available SYLGARD 184 silicone elastomer kit. To prepare the PDMS mixture, we thoroughly mix Silicone Elastomer and curing agent with a weight ratio of 10:1 followed by a debubbling process under rough vacuum. This mixture is spin coated on a silicon wafer with a spin rate of 350 rpm for 30 s, then cured at 80 °C for 30 min on a hot plate. The resultant PDMS is easily peeled off the silicon wafer for use.

<u>First-principles calculations.</u> We use the generalized gradient approximation (GGA)[36] and the projector augmented wave potentials (PAW)[37] as implemented in VASP.[38,39] The Grimme-D3 correction scheme is used to account for van-der-Waals interactions when performing structural optimization for the different stacking configurations.[40] The in-plane lattice constant is 3.293 Å for the 3R (AB and BA stacking) bilayer structure and is 3.292 Å for the 2H (ABBA) structure. Forces are converged to < 1 meV/Angstrom for each structure. We use a vacuum spacing of around 20 Å along the *c*-axis to avoid spurious interaction with periodically repeated cells. We used W PAW potentials with 5p, 6s and 5d valence electrons and Se PAW potentials with 4s and 4p valence electrons. We use an 8x8x1 *k*-point grid with a 450 eV energy cutoff for relaxation and for self-consistent electronic calculations. Spin orbit coupling is not included when optimizing the structures but is included in all electronic calculations.

The zone-center phonons for each of the optimized structures were calculated using density functional perturbation theory (DFPT) with a 28x28x1 *k*-point grid and the cutoff was increased to 600 eV.

The efficiency of a Raman process for scattering by an optical phonon is proportional to the derivative of the dielectric tensor elements with respect to the normal coordinate of the phonon, *Q*. In particular we calculate the cross, S($\omega$) which is defined as:

$$S(\omega) = \left|\frac{d\epsilon_i(\omega)}{dQ}\right|^2$$

Where $\omega$ is the frequency of the incoming photon, $\epsilon_i$ is the imaginary part of the frequency dependent dielectric function, and Q is the phonon coordinate of the normal mode which results in the displacement of the tungsten and/or chalcogen atoms.

We compute the efficiency of Raman processes using a frozen phonon approach. We use the eigenmodes of the $A_1$ and $E_2$ zone-center phonons of $WSe_2$ to determine the derivative of the in-plane dielectric function with the displacements of the normal phonon modes *Q*. The amplitude of the displacements were chosen to be small; the maximum atomic displacement was lower than 5% of the nearest neighbor W-Se bond length. The imaginary part of the dielectric function was calculated using a 28x28x1 *k*-point grid using tetrahedron integration with spin-orbit interaction included. Calculations using a 32x32x1 *k*-point grid lead to differences in the dielectric function on the order of 1 cm$^{-1}$. Excitonic interactions were not considered.

Conflicts of Interest:

There are no conflicts of interest to declare

Acknowledgements:

This work was supported by core programs at NRL and the NRL Nanoscience Institute. Computational work was supported by a grant of computer time from the DoD High Performance Computing Modernization Program at the U.S. Army Research Laboratory Supercomputing Resource Center.

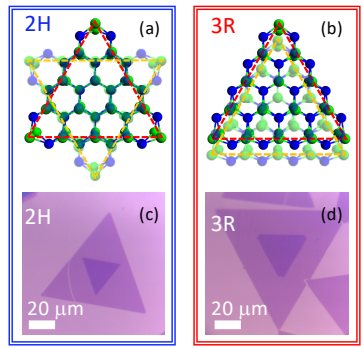

Figure 1. (a,b) Schematic drawings showing atomic positions for WSe$_2$ bilayers with 2H and 3R stacking order, respectively. Selenium atoms are depicted in green and tungsten in blue. Representative optical images of (c) 2H and (d) 3R WSe$_2$ bilayer samples.

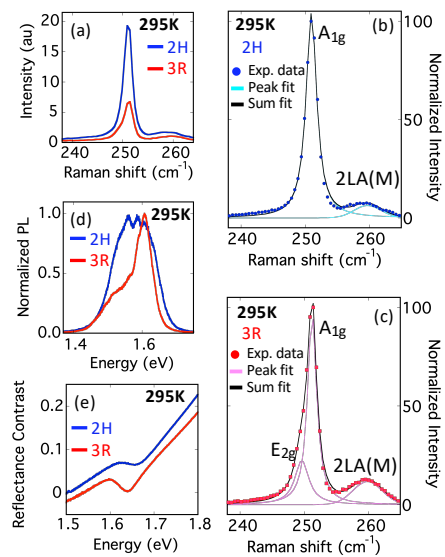

Figure 2. Room temperature Raman, PL and reflectance contrast spectra of bilayer WSe$_2$. 532 nm excitation is used for Raman and PL measurements (a) The Raman intensity of 2H stacking is a factor of 3 larger than for 3R stacking. (b) Bilayers with 2H stacking are well fit by a single Lorentzian at 251.0 cm$^{-1}$ and gaussian at 259.8 cm$^{-1}$, attributed to the A$_{1g}$ modes and second order 2LA(M) modes, respectively. (c) Bilayers of 3R stacking exhibit a low wavenumber shoulder, requiring two Lorentzian curves at 249.6 cm$^{-1}$ and 251.3 cm$^{-1}$ along with a gaussian at 259.8 cm$^{-1}$ to adequately fit the spectra, attributed to the E$_{2g}$, A$_{1g}$ and 2LA(M) modes, respectively. (d) Photoluminescence spectra exhibit clear differences in line shape. (e) In reflectance contrast spectra, the A-exciton is evident at 1.66 eV in 2H and 1.64 eV in 3R bilayers. All data are acquired after samples have been transferred to a fresh SiO$_2$/Si substrate to remove strain.

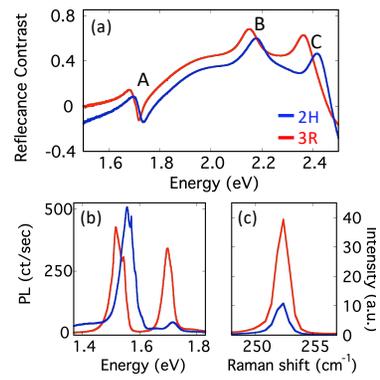

Figure 3. (a) Reflectance contrast, (b) PL, and (c) Raman of bilayer WSe$_2$ at 4K. 532 nm excitation is used for Raman and PL measurements. 2H orientation is displayed in blue and 3R in red.

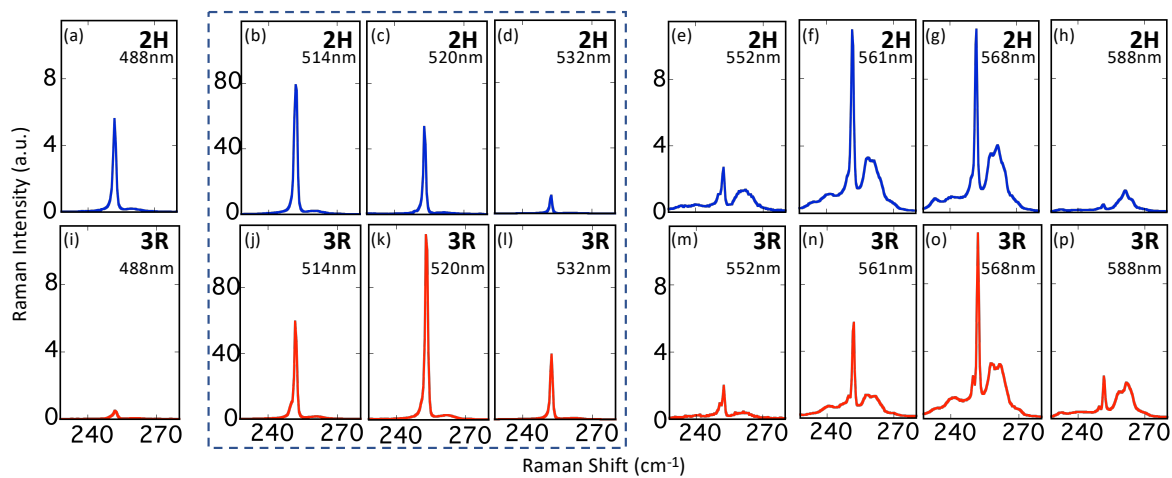

Figure 4. Low-temperature (4K) excitation-dependent Raman of WSe$_2$ bilayers for (a-h) 2H and (i-p) 3R stacking orientation. All spectra are normalized to Si Raman intensity at 523 cm$^{-1}$ and excitation wavelength is indicated in the upper right corner. Under 532 nm excitation, the A$_{1g}$ mode exhibits higher intensity in 3R than 2H samples, contrary to the room temperature behavior presented in figure 2a. Excitation from 514 nm to 532 nm leads to notably large A$_{1g}$ intensities in both 2H and 3R orientations, highlighted by the blue dashed box. However, in 2H samples the maximum intensity is evident for 514 nm excitation and for 3R maximum is evident under 520 nm excitation.

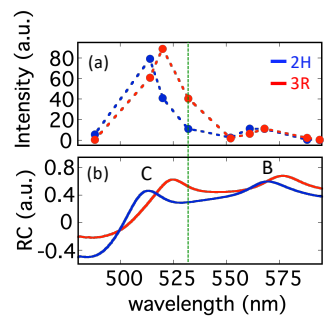

Figure 5. Low-temperature (a) excitation-dependent Raman intensity of the $A_{1g}$ mode and (b) reflectance contrast, RC, plotted as a function of wavelength. The Raman enhancements in 2H and 3R WSe$_2$ orientations coincide well with the C exciton feature in the reflectivity spectra.

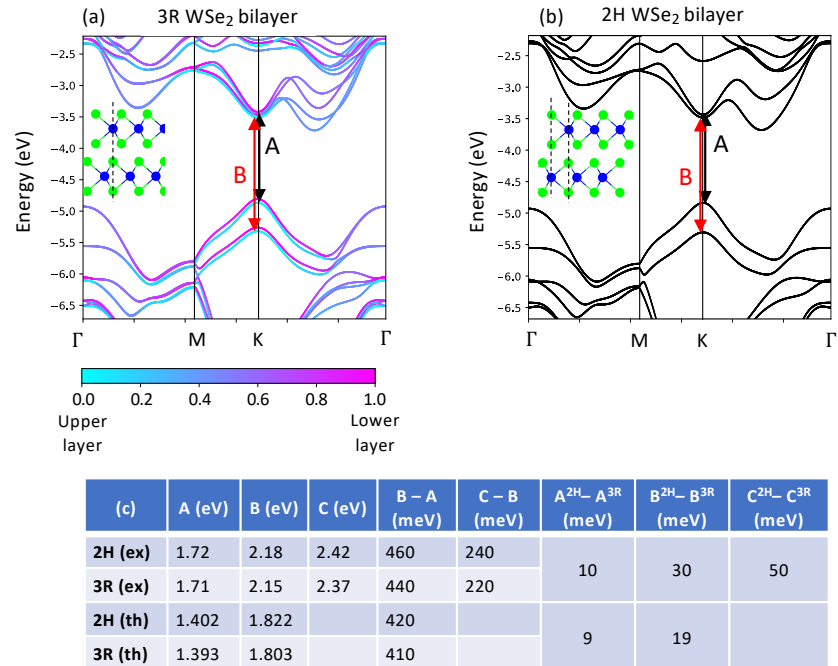

| (c) | A (eV) | B (eV) | C (eV) | B − A (meV) | C − B (meV) | $A^{2H}- A^{3R}$ (meV) | $B^{2H}- B^{3R}$ (meV) | $C^{2H}- C^{3R}$ (meV) |
|---|---|---|---|---|---|---|---|---|
| 2H (ex) | 1.72 | 2.18 | 2.42 | 460 | 240 | 10 | 30 | 50 |
| 3R (ex) | 1.71 | 2.15 | 2.37 | 440 | 220 | | | |
| 2H (th) | 1.402 | 1.822 | | 420 | | 9 | 19 | |
| 3R (th) | 1.393 | 1.803 | | 410 | | | | |

Figure 6. DFT calculations for (a) the 3R bilayer and (b) the 2H bilayer. Insets show the side view of a schematic of each stacking. Black (red) arrows show the origin in the band structure of the A and B excitons. All bands in the 2H bilayer are doubly degenerate, where each state in a degenerate pair may be layer-polarized or layer localized. (c) The table compares theoretical and experimental values for the band gaps. Although PBE-level theory underestimates band gaps, the differences in gap size agree well with differences in exciton energies measured in experiment.

# Supporting Information

## Stacking-Dependent Optical Properties in Bilayer WSe$_2$

Kathleen M. McCreary,[1] Madeleine Phillips,[1] Hsun-Jen Chuang,[2] Darshana Wickramaratne,[1] Matthew Rosenberger,[3] C. Stephen Hellberg[1] and Berend T. Jonker[1]

[1]*Naval Research Laboratory*, Washington, DC 20375
[2] *Nova Research, Inc.* Washington, DC 20375
[3] *University of Notre Dame*, Notre Dame IN 46556

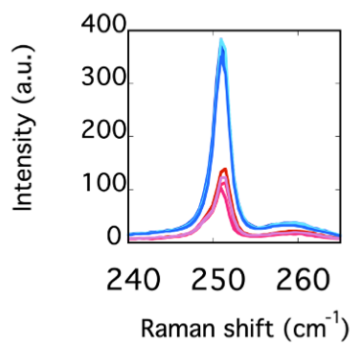

Figure S1. (a) The Raman spectra collected under 532nm excitation from four additional 2H bilayers (blue lines) and four additional 3R bilayers (red lines) confirm the behaviors observed in the main text, figure 2a. Bilayers with 2H stacking consistently exhibit higher intensity of the A$_{1g}$ mode.

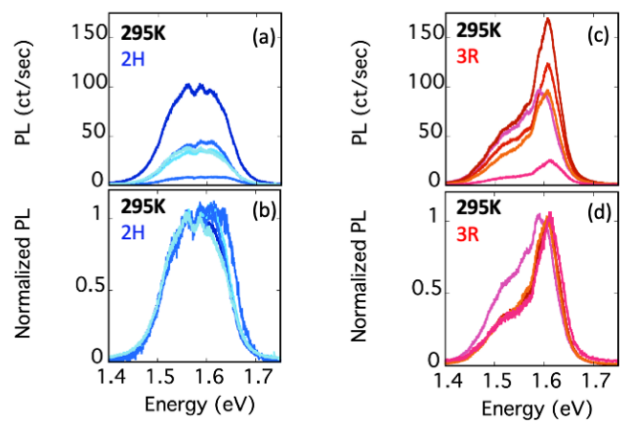

Figure S2. Room temperature photoluminescence from (a) five additional 2H samples and (c) five additional 3R samples. In both stacking orders the overall intensity varies significantly. (b,d) The normalized PL shows a consistent line shape for each stacking order, despite the intensity variations. 532nm excitation and identical parameters are used to probe all samples.

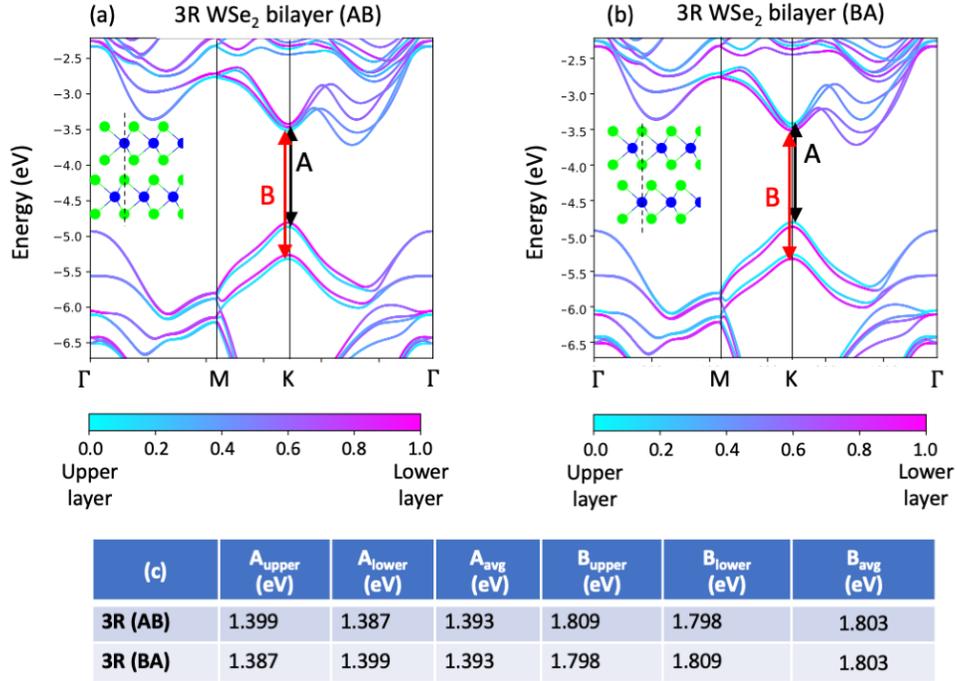

Figure S3. DFT calculations for 3R bilayers with (a) AB and (b) BA ordering. Insets show the side view of a schematic of each stacking. AB (BA) refers to a stacking geometry where W (Se) in the top layer is aligned with Se (W) in the bottom layer. The two stackings are related by a z → -z mirror operation. Black (red) arrows schematically show the origin in the band structure of the A (B) excitons. For each stacking there are two A excitons, one associated with the upper layer and one with the lower layer. Likewise there are two B excitons in each bilayer. (c) The table compares the band gaps associated with the A and B excitons in each stacking. The average of upper and lower A (B) bandgaps is the same in each stacking. These average values are quoted in the main manuscript.